\documentclass[useAMS,usenatbib, usegraphicx]{emulateapj}
\usepackage{aas_macros}
\usepackage{graphicx,epsfig} 
\usepackage{bbding}
\usepackage{wasysym}

\received{}
\revised{}
\accepted{Accepted for Publication}
\slugcomment{Accepted for Publication}
\shorttitle{Massive relic galaxies challenge the co-evolution of SMBHs and their host galaxies}
\shortauthors{A. Ferr\'e-Mateu et al.}

\begin{document}

\title{Massive relic galaxies challenge the co-evolution of super-massive black holes and their host galaxies}

\author{Anna Ferr\'e-Mateu$^{1}$, Mar Mezcua$^{2,3,4}$, Ignacio Trujillo$^{3,4}$, Marc Balcells$^{5}$ and Remco C. E. van den Bosch$^{6}$} 
\affil{$^{1}$Subaru Telescope, Hilo, HI 96720, USA}
\affil{$^{2}$Harvard-Smithsonian Center for Astrophysics, 60 Garden Street, Cambridge, MA 02138, USA}
\affil{$^{3}$Instituto de Astrof\'{\i}sica de Canarias, La Laguna, E38205 Tenerife, Spain}
\affil{$^{4}$Departamento de Astrof\'isica, Universidad de La Laguna, La Laguna, E38205 Tenerife, Spain}
\affil{$^{5}$ Isaac Newton Group of Telescopes, Santa Cruz de La Palma, E38700 Tenerife, Spain }
\affil{$^{6}$ Max-Planck Institute f\"ur Astronomie, 69117 Heidelberg, Germany }
\email{aferre@naoj.org (AFM)}   

\begin{abstract}

We study a sample of eight massive galaxies that are extreme outliers (3\,-\,5$\sigma$) in the M$_{\bullet}$\,-\,M$_\mathrm{bulge}$ local
scaling relation. Two of these galaxies are confirmed to host extremely large super massive black holes (SMBHs), whereas
the virial mass estimates for the other six are also consistent with having abnormally large SMBHs. From the analysis of
their star formation histories and their structural properties we find that all these extreme outliers can be considered
as relic galaxies from the early (z$\sim$2) Universe: i.e. they are compact (R$_{\mathrm{e}}$$<$2 kpc) and have purely old
stellar populations (t$\gtrsim$10 Gyr). In order to explain the nature of such deviations from the local relations, we
propose a scenario in which the hosts of these \textit{\"{u}ber-massive} SMBHs are galaxies that have followed a different
evolutionary path than the two-phase growth channel assumed for massive galaxies. Once the SMBH and the core of the galaxy
are formed at z$\sim$\,2, the galaxy skips the second phase, remaining structurally untouched and without further mass and
size increase. We show that if the outliers had followed the normal evolutionary path by growing in size via merger
activity, the expected (mild) growth in mass would place them closer to the observed local relations. Our results suggest
that the SMBH growth epoch for the most massive galaxies stopped $\sim$\,10\,Gyr ago.

 \end{abstract}


\section{Introduction}

It is now well established that the masses of super massive black holes (SMBHs; M$_{\bullet}>$\,10$^{6}$M$_{\odot}$)
strongly correlate with some of their host galaxies properties, such as their luminosities (e.g. \citealt{Kormendy1995};
\citealt{Gueltekin2009}; \citealt{Schulze2011}; \citealt{McConnell2013}; \citealt{Kormendy2013}), their masses (e.g
\citealt{Magorrian1998}; \citealt{Haering2004}; \citealt{Beifiori2012}), or their velocity dispersions ($\sigma$; e.g.
\citealt{Ferrarese2000}; \citealt{Gebhardt2000}; \citealt{Gueltekin2009}; \citealt{Graham2011}). This supports the idea
that the host galaxies and their SMBHs form and grow in a coordinated way by a common physical mechanism, which could be
AGN feedback, mergers or secular evolution (e.g. \citealt{Fabian1999}; \citealt{Hopkins2006}; \citealt{Somerville2008};
\citealt{Menci2008}; \citealt{Jahnke2011}). The correlation between SMBH mass and velocity dispersion is extremely tight,
following a log-normal distribution whose steepness increases as the galaxy sample is enlarged and more accurate
measurements of the SMBHs masses are used (see e.g. the most recent compilations of \citealt{McConnell2013} or
\citealt{Kormendy2013}). This correlation implies that the most massive galaxies
($\sigma\,\sim$\,300\,-\,400\,$\rm{km\,s^{-1}}$) should have SMBHs with M$_{\bullet}\sim$\,10$^{10}$M$_{\odot}$, which in
turn corresponds to high stellar masses (M$_\mathrm{bulge}\sim$\,10$^{12}$\,M$_\odot$) and high luminosities
(L$_\mathrm{V}\gtrsim$ 10$^{11}$ L$_\odot$).

However, in the last few years an increasing number of galaxies have been reported to strongly deviate from the above
scaling relations. These galaxies show larger SMBHs than what it is expected according to their velocity dispersions or
bulge masses. This is the case of e.g. the deeply analyzed NGC1277 (\citealt{vandenBosch2012}; \citealt{Emsellem2013};
\citealt{Fabian2013}; \citealt{Yildirim2015}), NGC4291 and NGC4342 \citep{Bogdan2012}, NGC1332 \citep{Rusli2011} and
SDSS\,J151741.75-004217.6 \citep[alias b19,][]{Laesker2013}.  All these extreme outliers lie far beyond the intrinsic
scatter of the relations, challenging the supposed co-evolution between SMBHs and their host galaxy ($\gtrsim$3$\sigma$). The
\textit{Hobby-Eberly Telescope Massive Galaxy Survey} (\textsc{hetmgs}; \citealt{vandenBosch2012,vandenBosch2015})
is a new large galaxy compilation aimed at searching nearby galaxies in which black hole masses can be directly measured. As a
by-product, they confirmed the existence of two other enormous SMBHs in NGC1271 \citep{Walsh2015} and MRK1216
\citep{Yildirim2015} and there are few dozens more candidates in their sample. Following \citet{Laesker2013} notation,
these type of extreme SMBHs have been named \textit{\"{u}ber-massive} SMBHs (\"UMBHs), although some critics have emerged
pointing out to an overestimation of the real mass. This overestimation could be due to the methodology employed (e.g.
long-slit spectroscopy instead of integral-field to derive proper dynamical models; e.g \citealt{Emsellem2013,
Yildirim2015}) or to variations in the initial mass function (which would vary the measured stellar mass of the host
galaxy; e.g. \citealt{Emsellem2013, Laesker2013}). Nonetheless, even accounting for these uncertainties, the galaxies
hosting \"UMBHs are still extreme outliers of the scaling relations. Consequently, it is worth addressing what are the
physical processes that cause these objects to be outliers. One possibility is that they are galaxies that did form the
corresponding stellar mass but then lost it by e.g. tidal stripping. However, the recent work from \citet{Bogdan2012}, who
studied two outliers in the low-mass regime, claimed that if that was the case, then the dark matter halo surrounding the
galaxy should also had been stripped. Their results, instead, showed that both galaxies reside in extended dark matter
haloes.

In this paper we propose and probe an alternative scenario. We hypothesize that galaxies hosting \"UMBHs form them in an
early phase of the galaxy evolution (z$\gtrsim$2) and once the SMBHs are formed, the galaxies remain structurally
untouched, which prevents them to reach the present-day scaling relations, leaving them as outliers. This early
formation is supported by the evolution of the M$_{\bullet}$\,-\,M$_\mathrm{bulge}$ scaling relation at high redshift
(e.g. \citealt{Walter2004}; \citealt{Jahnke2009}; \citealt{Greene2010}, \citealt{Petri2012}). Therefore, in our scenario,
massive galaxies hosting \"UMBHs are outliers in the black hole mass scaling relations because they did not follow the
expected two-phase growth channel, where merger accretion increases the size (and to less extent, the mass) of the host
galaxies after the central massive core is formed (e.g. \citealt{Naab2009}; \citealt{Oser2010}; \citealt{Hilz2013}). 
Therefore, we would expect the host of \"UMBHs to be galaxies showing the structural properties of the massive galaxy
population at z$\sim$2. In other words, they should be the ``relics'' of the early Universe, with old ($\gtrsim$10 Gyr) stellar
populations and compact (R$_{\mathrm{e}}\,\lesssim$2 kpc) sizes, similar to those found at high-z (e.g.
\citealt{Trujillo2007}; \citealt{Buitrago2008}; \citealt{Carrasco2010}). The existence of a small number of such galaxies
without merging accretion since z$\sim$2 is a natural prediction of the $\Lambda$CDM model (i.e. Quilis \& Trujillo 2013),
but to date only one candidate has been robustly confirmed \citep{Trujillo2014}. 

To probe such prediction, we explore the star formation histories (SFHs) for a sample of galaxies with \"UMBHs. If this
scenario is realistic, we should be able to pose a lower limit for the age of the SMBH formation, which corresponds to the
age of the host galaxy. The layout of this paper is the following. Section 2 describes the data used from the different
compilations and our spectroscopic sample. Section 3 analyses the derived SFHs of the sample in order to determine whether
a galaxy is a relic candidate or not. Section 4 shows the results and discusses their implications on the theories of the
co-evolution (or not) of the SMBHs and the host galaxy. A final summary is presented in Section 5. Throughout this work we
adopt a standard cosmological model with the following parameters: $H_0$=70 km s$^{-1}$ Mpc$^{-1}$, $\Omega_m$=0.3 and
$\Omega_\Lambda$=0.7.

\section{The data}

We first collect the largest possible sample of galaxies with SMBH mass measurements. There are three main recently
updated compilations of nearby galaxies: \citet[][McM13]{McConnell2013}, \citet[][KH13]{Kormendy2013} and
\citet[][GS13]{Graham2013}. They are all based on the initial sample of \citet{Gueltekin2009}, but also contain new
galaxies and improved measurements on the SMBH masses, galaxy distances and bulge mass determinations. All McM13 and GS13
galaxies (except for 9 new ones in the latter) are included in KH13. In addition, KH13 presents 13 new galaxies. This
literature compilation corresponds to a total of 95 galaxies.

A number of caveats are worth mentioning. First, galaxies found in common among the three compilations (McM13, KH13 and
GS13) do not always have the same reported values for some of the properties, as different techniques were employed.
Second, galaxies are morphologically classified following different criteria in each compilation. McM13 discriminate
between elliptical galaxies and bulges but do not differentiate between classical and pseudobulges, while KH13 make this
last separation on the bulges, and GS13 only separate by elliptical and disk galaxies. In addition, the authors of each
compilation do not include the same galaxies in their fits: some galaxies excluded in a fit may be included in another
one. Consequently, we use here the entire compilation to illustrate the correlations (Figure 1), but we are \textit{not}
deriving new fits for the correlations. In this exercise we use the fits corresponding to McM13 for all the galaxies with dynamical mass estimates, with
$\alpha$=\,8.32\,$\pm$\,0.05 and $\beta$=\,5.64\,$\pm$\,0.32 (for the M$_{\bullet}$\,-\,$\sigma$; Fig.\,1a) and
$\alpha$=\,8.46\,$\pm$\,0.08 and $\beta$=\,1.05\,$\pm$\,0.11 (for M$_{\bullet}$\,-\,M$_\mathrm{bulge}$; Fig.\,1b).

Together with the above compilation, we have expanded our sample using the new compilation of
galaxies from the \textsc{hetmgs} survey. This survey was  especially designed to find nearby
galaxies that were suitable for black hole mass measurements. It is composed of 1022 galaxies, hence is an ideal place where to find galaxies with very
large SMBHs, that will be the basis of our analysis for \"UMBH candidates, as explained in the next section.

\subsection{Defining the \"UMBH candidates}

\begin{figure*}
\includegraphics[scale=1.0]{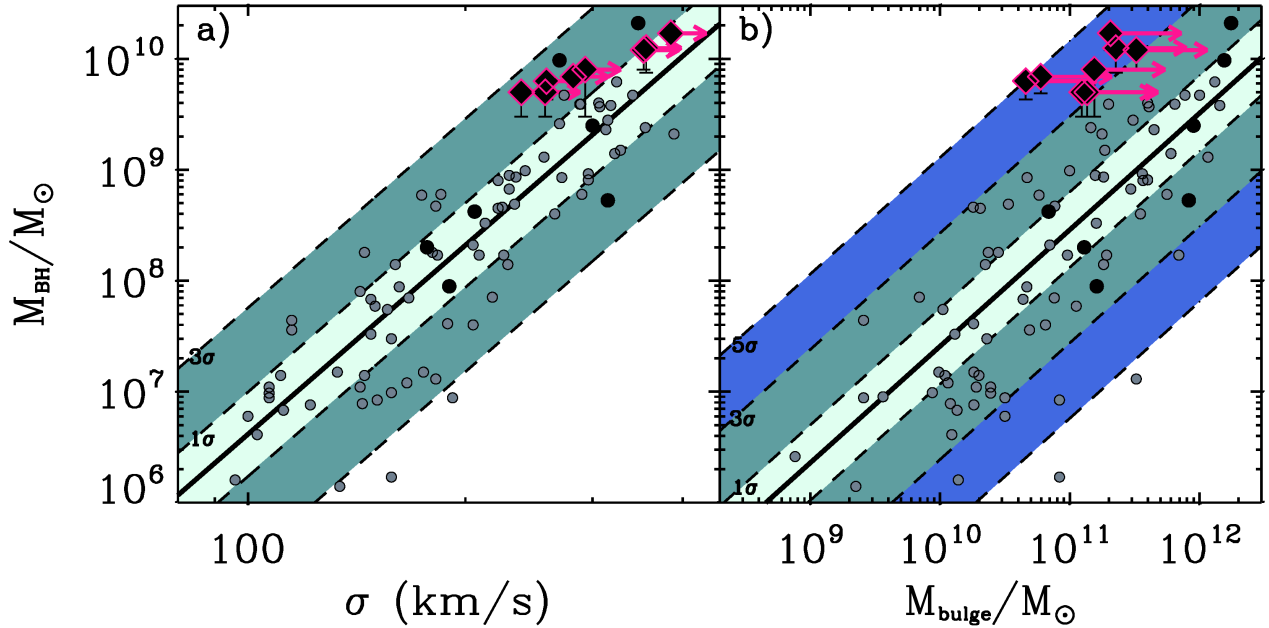}

\caption{Black hole mass scaling relations with velocity dispersion  (\textit{left}) and as function of bulge mass (\textit{right}).
Filled black diamonds correspond to the 8 \"UMBH host candidates, while circles correspond to the compilation of published galaxies with
SMBH detections (grey), with our 7 control galaxies emphasized in black large circles. The black solid line is the fit corresponding
to McM13 and the dashed lines limit the 1$\sigma$, 3$\sigma$ and 5$\sigma$ deviations from it.  Arrows mark the expected growth for
individual massive galaxies since z$\sim$2 if they had followed the merging phase of assembly for massive galaxies, making them more
normal in terms of SMBH mass (see section 4). The lower error bars show the $\sim$1.5 factor that the black hole mass is estimated
to possibly vary using more accurate techniques.}

\label{figure:1}
\end{figure*}

To identify \"UMBH host candidates, we select galaxies whose SMBHs are significantly larger than what is expected according to the
mass of the bulge of its host galaxy. In other words, we identify those galaxies which strongly deviate from the
M$_{\bullet}$\,-\,M$_\mathrm{bulge}$ scaling relation. We use the following operative definition to select \"UMBH host candidates: a
\"UMBH host candidate is a galaxy where the mass of its SMBH is located at a distance of 3$\sigma$ above  the
M$_{\bullet}$\,-\,M$_\mathrm{bulge}$ scaling relation given by  McM13 (see Fig. \ref{figure:1}). With this definition, it is very unlikely that the mass of its SMBH can be explained as simply produced by the uncertainty in the measurement of the
mass of its SMBH. To have a significant number of \"UMBH host candidates, we use the \textsc{hetmgs} sample. In this paper, we are interested in the most massive galaxies, hence we will concentrate our efforts on finding \"UMBH candidates in the upper right corner of the M$_{\bullet}$\,-\,M$_\mathrm{bulge}$ scaling relation.\\
From the \textsc{hetmgs} sample we first select those galaxies that are good candidates to host \"UMBHs by imposing
the following criteria:

\begin{itemize}
\item To have a large dynamical enclosed mass inside the central $R\,=\,3"$ aperture of the \textsc{hetmgs}, which
indicates a very high central mass concentration (from either stars or a black hole).  We use the virial mass estimator $M_{vir} =
\sigma_c^2\,R\,G^{-1}> >10^{9.7}$ M$_{\odot}$, where $G$ is the gravitational constant and $\sigma_c$ the central velocity
dispersion. This mass does not discriminate between baryonic mass and black hole mass. And as such, these virial masses are
considered upper limits for the black hole masses (see below).

\item To be nearby enough so that the putative black hole can be resolved with high spatial resolution facilities; i.e the
sphere of influence $\theta_i = G\,\sigma_c^{2.5}\,D^{-1}$ (being $D$ the Hubble flow distance) to be larger then $0.06"$ (see
\citealt{vandenBosch2015}).
\end{itemize}

With the above criteria, the sample of candidates from the \textsc{hetmgs} consists of 173 galaxies. These galaxies
are, as expected, average massive galaxies with sizes of $\sim$4\,kpc. The vast majority of them do not qualify as \"UMBH host
candidates (i.e. they are not 3$\sigma$ outliers of the M$_{\bullet}$\,-\,M$_\mathrm{bulge}$ scaling relation). With the 3$\sigma$
criteria, our sample is reduced to 30 \"UMBH candidates. Interestingly, this last criteria changes dramatically the structural properties of the selected galaxies: they have now an average size of $\sim$1.5 kpc. In other words, the \"UMBH host candidates are massive compact galaxies. The sizes of the
\textsc{hetmgs} galaxies have been retrieved from the SDSS database as the half-light effective radii in the $r$-band
($r_{e}$) and then circularized (R$_{e}$).\\
We need to make a last selecting criteria, which is to have pre-existing SDSS\footnote{Sloan Digital Sky Survey} spectroscopy with
high quality (S/N$\ge$\,20). This is crucial to robustly derive the SFHs of the galaxies with the full-spectral-fitting technique
employed in Section 3. This reduces our sample of  \"UMBH candidates to 10 galaxies. Two of them (NGC0426 and NGC2522) were rejected
as their spectra were classified by SDSS as Active Galaxy Nucleus (AGN) broadline. Having an AGN would not only make the
determination of the stellar population properties of the galaxies uncertain but also prevent a proper estimation of the structural
properties of the galaxies. In fact, the presence of a bright nucleus could bias the measurement of the effective radius of the
galaxies towards smaller sizes. In order to avoid this potential source of contamination from our sample, we have ended with a final
sample of 8 \"UMBH candidates.

We use the central enclosed virial mass estimate from the \textsc{hetmgs} as an estimation of the M$_{\bullet}$ of the host
galaxies. This means that both  the black hole masses and their locus in the SMBH-host galaxy planes should be considered as an
upper limit. Efforts towards obtaining more accurate estimates using orbit-based dynamical models are underway. For example,
the actual black hole mass for NGC1277 has been lowered to 12$\times\,10^{9}$ M$_\odot$ \citep{Yildirim2015}, which is a factor of
1.5 lower than the upper limits. Therefore, if we apply this factor to the rest of the \"UMBH candidates we have a lower, more
conservative, estimate of their black hole mass. Nonetheless, in the absence of high-resolution spectroscopy and orbit-based models
for our sample of candidates, we will use the upper limit estimates for our analysis. However, it is worth noticing that, even
considering the lower estimates, the \"UMBH candidates are still extreme outliers in the local M$_{\bullet}$\,-\,M$_\mathrm{bulge}$
scaling relation (Fig.\,1b; $\sim$\,3\,-\,5$\sigma$ deviations). It is worth noting that, when placed in the
M$_{\bullet}$\,-\,$\sigma$ scaling relation, our  \"UMBH host candidates are not that extreme (Fig.\,1a; $\sim$\,1\,-\,3$\sigma$ deviations). The reason for such
differences between the two planes is further discussed in Sect.\, 4. The details of our complete spectroscopic sample are described
in Table 1.

\begin{table*}
\label{table:1}                      
\centering
\caption{The spectroscopic sample}    
\begin{tabular}{c|c c c c c c c}   
\hline   
\textbf{Galaxy} & Spectra & M$_{\bullet} (+,-)$ & M$_{\mathrm{bulge}} (+,-)$  & vel. disp. & R$_{\mathrm{e}}$ & deviation $(+,-)$ & deviation $(+,-)$ \\  
& & (10$^{9}$ M$_\odot$) & (10$^{11}$ M$_\odot$) & (km$\,\mathrm{s^{-1}}$) & (kpc) &  M$_{\mathrm{bulge}}$ & vel. disp.\\
\hline  \hline
\multicolumn{8}{c}{\textbf{\"UMBH host candidates from} \textsc{hetmgs}}\\
\hline
NGC1270   &  SDSS & $<$\,12  & 3.3 (2.7,1.3) & 353$\pm$\,4 & 1.92$\pm$\,0.03  & $3.2\sigma$ (-,1.6)&  $1.0\sigma$ (-,0.7)\\     
NGC1271   &  SDSS & $<$\,8    & 1.5 (0.8,0.6) & 293$\pm$\,6 & 1.31$\pm$\,0.02  & $4.0\sigma$ (-,2.0)&  $2.0\sigma$ (-,1.9)\\    
NGC1277   &  SDSS & $<$\,17  & 2.1 (1.9,0.7) & 385$\pm$\,4 & 1.64$\pm$\,0.03  & $4.2\sigma$ (-,1.4)&  $0.8\sigma$ (-,0.7)\\    
NGC1281   &  SDSS & $<$\,5    & 1.2 (1.0,0.4) & 258$\pm$\,4 & 1.46$\pm$\,0.03  & $3.4\sigma$ (-,0.6)&  $2.0\sigma$ (-,0.9)\\    
NGC2767   &  SDSS & $<$\,5    & 1.4 (1.1,0.6) & 239$\pm$\,3 & 1.95$\pm$\,0.03  & $3.2\sigma$ (-,1.7)&  $2.5\sigma$ (-,1.3)\\ 
PGC012557 &  SDSS & $<$\,7   & 0.6 (0.6,0.3) & 283$\pm$\,4 & 0.71$\pm$\,0.02  & $4.7\sigma$ (-,1.0)&  $1.8\sigma$ (-,0.8)\\      
PGC012562 &  SDSS & $<$\,6   & 0.4 (0.3,0.2) & 259$\pm$\,4 & 0.70$\pm$\,0.02  & $5.0\sigma$ (-,1.2)&  $2.2\sigma$ (-,0.7)\\   
PGC032873 &  SDSS & $<$\,14 & 2.3 (1.2,0.9) & 356$\pm$\,3 & 1.76$\pm$\,0.03  & $3.7\sigma$ (-,1.2)&  $1.0\sigma$ (-,0.8)\\   
\hline
\multicolumn{8}{c}{\textbf{Control sample from McM13}}\\
\hline
NGC3379  &  Y06    & 0.4 (0.1,0.1)  & 0.7 (0.5,0.3)    & 206$\pm$\,10  & 3.3$\pm$\,0.4   &  $1.0\sigma$ (1.1,1.1)&  $0.6\sigma$ (0.6,0.8)\\  
NGC3842  &  SDSS & 9.7 (3.0,2.5)  &16.0 (14.0,7.0) & 270$\pm$\,14  & 14.0$\pm$\,1.3 &  $0.8\sigma$ (1.1,1.2)&  $2.4\sigma$ (0.7,0.9)\\  
NGC4261  &  SDSS & 0.5 (0.1,0.1)  & 8.3 (7.0,4.0)    & 315$\pm$\,15  & 7.4$\pm$\,0.7   &  $2.1\sigma$ (1.1,1.1)&  $1.9\sigma$ (0.5,0.8)\\  
NGC4472  &  Y06    & 2.5 (0.6,0.1)  & 9.0 (7.0,4.0)    & 300$\pm$\,15  & 8.2$\pm$\,0.4   &  $0.2\sigma$ (0.8,1.1)&  $0.2\sigma$ (0.6,0.7)\\  
NGC4473  &  Y06    & 0.1 (0.0,0.0)  & 1.6 (2.0,0.6)    & 190$\pm$\,9    & 4.3$\pm$\,0.3   &  $2.1\sigma$ (0.9,1.6)&  $0.6\sigma$ (0.7,1.0)\\ 
NGC4697  &  Y06    & 0.2 (0.0,0.0)  & 1.3 (0.7,0.6)    & 177$\pm$\,8    & 6.4$\pm$\,0.4   &  $0.8\sigma$ (1.3,1.5)&  $0.7\sigma$ (0.8,1.5)\\  
NGC4889  &  SDSS & 21.0 (16.0,15.5) & 18.0 (14,8) & 347$\pm$\,17  & 12.6$\pm$\,1.4 &  $1.6\sigma$ (1.6,2.5)&  $1.7\sigma$ (1.0,2.6)\\                  
\hline                                  
\end{tabular}

{Column 1 -- Our spectroscopic sample from the compilation of \textsc{hetmgs} (top) and McM13 (bottom). Column 2 -- origin of the
spectra: SDSS or Y06 (from Yamada et al. 2006). Columns 3 and 4 -- the masses of the SMBHs and the masses of the bulge of the galaxy
from their dynamical estimates (from \textsc{hetmgs} and McM13). Column 5 -- Velocity dispersions (derived with {\tt STARLIGHT} and
from McM13). Column 6  -- Circularized effective radii (from the SDSS $r$-band, ATLAS3D and Hyperleda). Columns 7 and 8 --
Deviations from the local SMBH mass scaling relations. Note that for the \"UMBH host candidates the black hole masses are an upper limit,
thus marked with an $<$.}

\end{table*}

\subsection{The control sample}

In order to test our scenario, we also explore the ages of the stellar populations of galaxies in a control sample. They
were selected from the above (McM13, KH13, GS13) compilations uniquely by the criteria of having a SMBH detection and having
high-quality, flux calibrated spectra available. No selection based upon morphology, size or any other parameter was done. Only 3
galaxies out of the 95 have good quality SDSS spectra (NGC3842, NGC4261 and NGC4889). High quality archive spectroscopic data is
also included for a few other control galaxies: NGC3379, NGC4472, NGC4473 and NGC4697 (from \citealt{Yamada2006}, the reader is
referred to this publication for a further description of the data). Their black hole masses, velocity dispersions and bulge masses
are those quoted in Table A1 from McM13. The circularized sizes have been retrieved from \citet{Krajnovic2013} for those galaxies
belonging to the \textsc{ATLAS3D} survey (\citealt{Cappellari2011}; NGC3379, NGC4261, NGC4472, NGC4473 and NGC4697) and
from Hyperleda\footnote{http://leda.univ-lyon1.fr} for NGC3842 and NGC4889. This control sample is thus composed of large
ellipticals and S0s mostly from Coma and Virgo clusters. They cover a larger parameter space than our \"UMBH candidates and are
typically within the intrinsic scatter of the local relations, showing smaller deviations.

\section{Analysis}

As mentioned before, under the proposed scenario present-day massive galaxies hosting \"UMBHs should have the
characteristics of the typical population of galaxies at z$\sim$2 after passively evolving since that epoch. In other
words, they should be massive (M*$\gtrsim$\,10$^{11}$ M$_\odot$), compact (R$_{\mathrm{e}}\,\lesssim$2 kpc) and with old
stellar populations ($\gtrsim$10 Gyr). Galaxies that fulfill these criteria in the local Universe are named
\textit{relic galaxies} and are known to be extreme outliers from the local mass-size relation \citep{Trujillo2009}. 
Therefore, we have quantified the compactness of our galaxies by measuring how much their size deviates from the expected
value according to the present-day local size-mass relation of \citet{Shen2003}. This well-known relation was derived
using a complete sample of 140,000 galaxies from the SDSS. The stellar masses used in that paper were taken from \citet{Kauffmann2003},
calculated by multiplying the SDSS Petrosian luminosity with the model-derived mass-to-light ratio. The radii considered
were the S\'ersic half-light radius $R_{50,S}$ in the $z$-band. For galaxies with $n\,>$\,2.5, the local mass-size
relation is:

\begin{equation}
\bar{R}\,(kpc)\,=\,b\,(M/M_{\odot})^{a}
\end{equation}

where $\bar{R}$ is the mean radius, $M$ is the stellar mass, and $a$ and $b$ are fitting parameters given by a
least-squares fit to the data, $a$\,= 0.56 and $b$=\,2.88$\times$\,10$^{-6}$. This relation is shown as the solid green
line in Figure 2. This figure indicates the location of our spectroscopic sample in the mass-size relation plane. It is
clear that our \"UMBH candidates largely deviate from the local relation.

We consider here that a galaxy is compact if the ratio R$_{\mathrm{e}}$/R$_{\mathrm{e,shen03}}\,\lesssim$0.33 (see Table 2). This corresponds to
$\sim$\,1/3 of the normal size that massive galaxies show at z\,=\,0 at a given stellar mass. The location in the stellar mass-size
plane of our \"UMBH host candidates is equivalent to the locus occupied by the massive compact galaxies at high-z.

\begin{figure}
\centering
\includegraphics[scale=0.47]{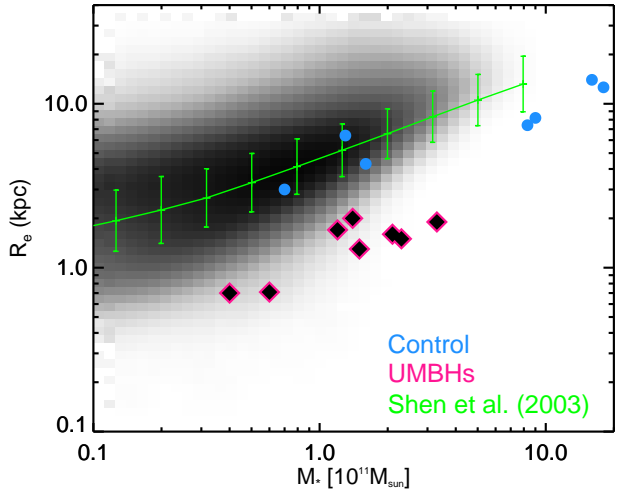}\\

\caption{Stellar mass-size distribution of the SDSS galaxies. Overplotted in green, the Shen et al. (2003) mass-size
distribution for SDSS $n>$2.5 galaxies is shown. The position of our \"UMBH host candidates is shown with black diamonds, while the
control sample is shown in blue filled dots. It is clear that all our \"UMBH host candidates are extreme outliers of the local
mass-size relation and thus good candidates to be massive relics of the early Universe.}

\label{figure:2}
\end{figure}

To determine the stellar populations of both the outliers and the control sample we derive the SFHs of our galaxies
applying a full-spectral-fitting approach. This technique has several advantages compared to the classical line strength
approach. One of them is that the results are not limited to luminosity-weighted estimates but also give the output in
terms of mass, allowing to recover the true fossil imprints of the epoch when the stars were created. The code employed
here is {\tt STARLIGHT} (\citealt{CidFernandes2005}; \citealt{CidFernandes2010}), which creates a combination of single
stellar populations (SSPs) model predictions that best resemble the observed galaxy spectrum while minimizing the
$\chi^{2}$. The reader is referred to the source code papers for a deeper description of the methodology, or to
the Appendix for a brief summary of this technique. The SSP models used are {\tt MIUSCAT} (\citealt{Vazdekis2012};
\citealt{Ricciardelli2012}), which cover a wide range of both ages (0.1-17.8\,Gyr) and metallicities (-1.71 to 0.22\,dex).
For this exercise in particular, 46 ages and 4 different metallicites were considered, making a total of 184 SSP
templates. These SSP models also allow for variations on the IMF slope and shape. The IMF has been recently reported to
be non-universal (e.g. \citealt{Cappellari2012}; \citealt{Conroy2012a}; \citealt{Spiniello2012}; \citealt{Ferreras2013};
\citealt{LaBarbera2013}) and to have a strong impact in the derived stellar population properties \citep{Ferre-Mateu2013}.
However, \citet{Martin-Navarro2015} have shown recently that for the compact massive relic galaxy NGC1277, changing the
IMF does no modify the derived age properties. This indicates that it is still unclear which is the parameter that really
drives these IMF variations (e.g. velocity dispersion, metallicity, $\alpha$-enhancement). Therefore, here we will focus
on the standard assumption of a Kroupa Universal IMF. 

Figure\,3 illustrates the derived SFH from the spectral fitting for one of the galaxies (NGC1271). The derived SFHs for
the rest of our spectroscopic sample can be found in the Appendix. The right panels in Figure\,3 show the derived SFHs as
the percentage of mass created at a given age, considering both the contribution of each individual metallicity from the
models and of the total metallicity. It is worth emphasizing here an important caveat on the form the derived SFHs are
shown. As mentioned by \citet{CidFernandes2005}, the reader should not take the form of \textit{each single burst} at face
value. This is partly because the SSP models do not cover uniformly the entire age parameter space. For example, there is
always a gap at 13\,Gyr because there is no SSP model corresponding to that age in the {\tt MIUSCAT} models. Therefore,
the derived SFHs should be better interpreted as clearly differentiated epochs of formation, i.e. young, intermediate and
old. These considerations are fundamental for the purpose of our exercise, as one of our criteria to consider a galaxy as
a relic of the z$\sim$2 Universe is that all their stellar populations must be old ($>$10\,Gyr). When looking at the age
distributions, it can be seen that sometimes there is a small contribution at 8\,-\,9\,Gyr, which can be considered as the
tail of the distribution that peaks at $\sim$\,10\,Gyr. For this reason, in order to be more quantitative in our statement
whether a galaxy has or not entirely old stellar populations, we measure the fraction of mass $\Omega$ which is within the
interval 8$<$t$<$10 Gyr. We thus consider a galaxy compatible with having all its stellar mass formed at z$\gtrsim$\,2 if
there are no stellar populations younger than 8\,Gyr and $\Omega$ is below 10$\%$ of the total mass.

We summarize in Table 2 the quality parameters from the fit and whether the outliers satisfy our criteria to be considered
relic galaxies or not. We find that 7 out of the 8 objects undoubtedly qualify as relic galaxies, while only one is
dubious. As these compact relic galaxies are also outliers in the present-day mass-size relation (Figure 2), finding these
\"UMBH host candidates could constitute a new way of detecting these rather elusive relic galaxies in the nearby Universe
(e.g. \citealt{Trujillo2009}; \citealt{Taylor2010}; \citealt{Ferre-Mateu2012}; \citealt{Trujillo2012},
\citealt{Poggianti2013}). Regarding our control sample, none of the galaxies can be considered as massive relic of the
high-z Universe because of their large sizes. However, a couple of them are as old as our relic
galaxies, indicating that at least the stellar populations of their most central parts were formed at very early epochs too. We would like to
emphasize that our criteria based on the $\Omega$ value, in order to be fully consistent with the statement that the
galaxy is a relic, should be shown to be true \textit{along the entire structure} of the galaxy. As we are employing SDSS
data, the coverage of the fiber is $\sim$\,0.3\,-\,0.8R$_{e}$ for the compact galaxies and $\sim$0.01-0.03R$_{e}$ for the
control sample. Consequently, the SDSS data does not allow us to perform such radial study. However, this issue has been assessed for the
compact candidates in \citet{Trujillo2014} for NGC1277 and in the companion paper of Ferr\'e-Mateu (2015b, in prep.) for
PGC032873. These works both use high-quality spectroscopy from the William Herschel Telescope to the derive SFHs and
stellar populations properties out to several effective radii, showing no gradients and supporting our hypothesis of this
type of galaxies being relics.

\begin{figure*}
\centering
\includegraphics[scale=0.9]{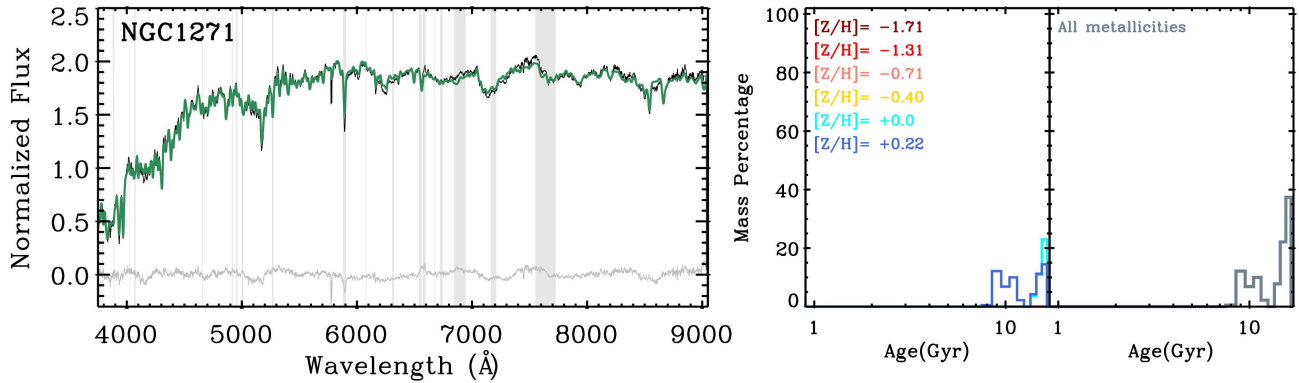}\\
\caption{Illustration of the full-spectral-fitting approach used in this work for one of our \"UMBH host candidates. The whole sample can be found in the Appendix. The left panel shows the galaxy NGC1271 SDSS spectrum (black), the fitting from {\tt STARLIGHT} (green), and the residuals from the fitting and the masked spectral regions to avoid possible emission line contributions (grey). The right panel shows the mass-weighted derived SFHs considering both the contribution from each individual metallicity and from the total metallicity.}
\label{figure:3}
\end{figure*}

\begin{table*}
\label{table:2}                      
\centering
\caption{Fitting and Massive galaxy relic criteria}    
\begin{tabular}{c| c c |c c c c}   
\hline     
\textbf{Galaxy} & [$\chi^{2}$] & adev\,($\%$) & R$_{\mathrm{e}}/\mathrm{R_{e,shen03}}$ & $\%\mathrm{M(t\,<\,8}$Gyr) & $\Omega(\%)$ & \textbf{Relic?} \\  
\hline\hline
\multicolumn{7}{c}{\textbf{\"UMBH host candidates}}\\
\hline  
NGC1270     & 1.6 & 1.7 & 0.24   & 0 & 7   & \checkmark  \\  
NGC1271     & 1.7 & 1.6 & 0.25   & 0 & 10 & \checkmark   \\    
NGC1277     & 1.9 & 1.5 & 0.26   & 0 & 0   & \checkmark  \\      
NGC1281     & 1.2 & 1.6 & 0.33   & 0 & 0   & \checkmark  \\    
NGC2767     & 0.7 & 1.8 & 0.37   & 0 & 0   &  \checkmark  \\
PGC012557 & 1.3 & 1.7 & 0.22    & 0 & 0  &  \checkmark   \\     
PGC012562 & 1.3 & 1.3 & 0.23   & 0 & 14 & $\sim$ \\	 
PGC032873 & 1.0 & 1.0 & 0.23   & 0 & 0  &  \checkmark  \\     
\hline
\multicolumn{7}{c}{\textbf{Control}}\\
\hline  
NGC3379   & 0.1 & 1.9 & 0.92  & 5  & 4   & \XSolidBrush \\    
NGC3842   & 0.8 & 2.0 & 0.73  & 1  & 14 & \XSolidBrush \\    
NGC4261   & 0.7 & 1.8 & 0.55  & 0  & 5   & \XSolidBrush \\    
NGC4472   & 1.5 & 1.7 & 0.55  & 7  & 5   & \XSolidBrush \\    
NGC4473   & 1.1 & 1.3 & 0.83  & 1  & 14 & \XSolidBrush \\    
NGC4697   & 1.2 & 1.4 & 1.26  & 21& 7   & \XSolidBrush \\    
NGC4889   & 0.8 & 1.8 & 0.61  & 0  & 10 & \XSolidBrush\\			  
\hline                                  
\end{tabular}

{The table summarizes the different criteria by which a galaxy can be considered a relic candidate.  Column 2 and 3 are measurements of the quality of the fit from {\tt STARLIGHT}, (see Appendix). Column 4 measures the compactness of a galaxy, by showing how much the galaxy size deviates from the present-day mass-size relation of \citet{Shen2003} for $n\,>$\,2.5 galaxies. Column 5 and Column 6 quote the fraction of stellar populations with $t<8$\,Gyr, and the fraction of mass with 8$<$t$<$10 Gyr, $\Omega$. Column 7 states the verdict on whether the galaxy is considered a relic candidate or not.}
\end{table*}

\section{Discussion}
The tight correlations that the mass of the SMBHs and their host galaxy properties follow have been considered as a proof of the co-evolution between galactic bulges and the SMBHs they host. According to this paradigm, they should be coupled by a common physical mechanism, e.g. feedback, mergers, secular evolution. However, a small sample of galaxies hosting \"UMBHs now challenge this assumed universal co-evolution as they clearly fail to follow such relations. In this paper we have proposed that the nature of these outliers hosting \"UMBHs is connected to the uncommon evolutionary path followed by these massive galaxies (without growing in mass or size). It is generally assumed from galaxy formation models that massive galaxies form in a two-phase mechanism, where the central massive part of the galaxy is created in a fast and violent event at high-z, and then the galaxy grows by posterior merger activity (e.g. \citealt{Naab2009}; \citealt{Oser2010}; \citealt{Hilz2013}). But because mergers are stochastic events, it is expected that some galaxies remain untouched over cosmic time, remaining compact and entirely old (see \citealt{Quilis2013}). Therefore, if the \"UMBH-outliers are those relics, as our results suggest, then the nature of the deviations could be explained by the lack of merging activity and its consequent lack of galaxy growth. Under our hypothesis, these galaxies should occupy in the SMBH mass scaling relations the position of the population of galaxies at z$\sim$2. Consequently, present day most massive galaxies should start at those locations $\sim$10 Gyr ago and their growth in mass and size should be able to move them to the current scaling relations. If this is true, the expected mass and $\sigma$ increase of the massive galaxies since z$\sim$2 should be enough to locate our outliers in agreement with the local relations. 

We explore this possibility on what follows. The arrows in Figure 1 show the estimated growth in velocity dispersion and stellar mass since z$\sim$\,2 for \textit{individual} massive galaxies. From semi-analytical models, if an individual galaxy grows $\sim$7 times in size, it can increase its mass by almost a factor of 5 (e.g. \citealt{Oser2010}; \citealt{Trujillo2011}). This could account for the missing stellar mass, placing the \"UMBH-outliers closer to the central distribution in the M$_{\bullet}$-M$_\mathrm{bulge}$ relation. As commented above, our sample is not that extreme when placed in the M$_{\bullet}$\,-\,$\sigma$ relation, hence no much variation in the velocity dispersion should be expected, if any. This is in agreement with the model predictions (e.g. \citealt{Cenarro2009}; \citealt{Oser2012}; \citealt{Oogi2013}; \citealt{Wellons2015}), where the velocity dispersion varies very slightly since z$\sim$\,2. Depending on the dominant merger channel, individual galaxies can increase very mildly their velocity dispersion by a factor of $\sim$1.1 (\citealt{Hilz2012}; Tapia et al. 2015 subm.). It is worth noting that in our scenario, the arrows move horizontally because an increase in the SMBH is expected to be almost negligible during the second phase. It is typically assumed that the amount of mergers that occur since z$\sim$\,2 have a mass ratio of 1:3 to 1:5 (\citealt{Oser2010, Oser2012}). Here we assume that the massive galaxies, once formed at z$\sim$\,2, have a much larger SMBH than what they should have from their stellar masses. Then, they merge with smaller satellites, whose mass-size evolution is milder over cosmic time \citep{Trujillo2004} and therefore are expected to be located already in, or very close to, the local relation having smaller black holes. This type of encounters would produce a negligible increase in the mass of the SMBH, as seen in cosmological simulations of SMBH growth (e.g. \citealt {Yoo2007, Wellons2015, Kulier2015}). Therefore, it is a valid assumption to account for evolution only in the M$_{\mathrm{bulge}}$ and $\sigma$ directions for the relic galaxies and neglect the one in the SMBH mass direction. As a final comment, considering the lower black hole estimates would further support our scenario: the  \"UMBH candidates would still be extreme outliers in the M$_{\bullet}$-M$_\mathrm{bulge}$ relation but they would be located even closer to the local scaling relations after accounting for the missing mass.
  
In a nutshell, our scenario is schematically depicted in Figure\,4. The left panel shows the normal evolution that a massive galaxy follows over cosmic time to become a large elliptical galaxy today, with the two-phase of formation \citep{Oser2010}. The right panel illustrates the expected evolutionary track for the relic/outlier galaxies. We remind the reader that this is a cartoon to illustrate the proposed scenario, where a number of assumptions have been here simplified, as many questions in the (co-)evolution of SMBHs and their host galaxies remain unsolved. For example, the exact position in the early Universe phase. It is not yet clear if the SMBH and the host started forming at the same time, but several observational and theoretical works suggest that the SMBH started forming first, as inferred from the mild deviations of these objects in the M$_{\bullet}$-$\sigma$ (e.g. \citealt{Walter2004}; \citealt{Jahnke2009}, \citealt{Merloni2010}; \citealt{Reines2011}; \citealt{Petri2012}; \citealt{Khandai2012}). Our scenario also seems to support this case, as by the end of the z$\gtrsim$\,2 phase the SMBH should be almost fully in place, being larger than expected from the galaxy stellar mass.  Although we are not in the position to confirm this from our derived SFHs, we can pose a lower limit for the SMBHs growth at $\sim$\,10\,Gyr, when the star formation activity in the host galaxy is halted. At this point, one would expect the massive compact galaxy to move into the second phase of formation, balancing out the deviations with its SMBH, which roughly evolves, by growing largely in size and mildly in mass through mergers. This will eventually place the massive galaxy in the local M$_{\bullet}$\,-\,M$_\mathrm{bulge}$ relation. 

We are aware of the limited sample employed here, but so is the published sample of galaxies with detected SMBHs and accurate measurements or the sample of relic galaxy candidates. In a recent effort, \citet{Saulder2015} have revisited the number of relic candidates in the SDSS, enlarging the one from \citet{Trujillo2009} to 76 candidates. The next step is to find out how many of these galaxies host a SMBH and if so, if they are \"UMBHs. This will be achieved with the upcoming era of the largest telescopes, when SMBH detections and measurements will be improved. On the meantime, surveys such as the \textsc{hetmgs} are ideal compilations for new SMBH detections. A deeper understanding on the formation of these extreme outliers (both in the upper and the low-mass end) is crucial to test the current theories of galaxy formation and evolution and cosmological models. 

\begin{figure*}
\centering
\includegraphics[scale=0.36]{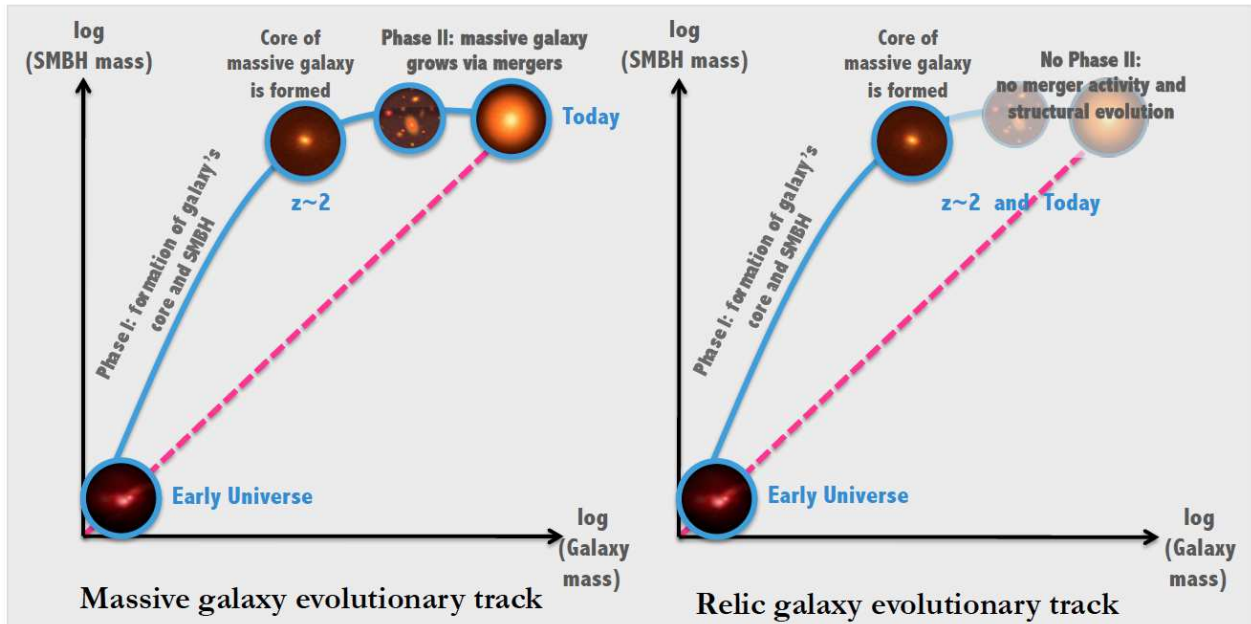}
\caption{Schematic view of the scenario proposed in this work: galaxies hosting \"UMBHs depart from the mass scaling relationship due to their unusual growth channel. The figure shows the assumed evolutionary track over cosmic time in the SMBH mass- host galaxy mass plane for massive galaxies following the two-phase growth channel (\textit{left}) and for the relic galaxies (\textit{right}), who skip the second phase of such channel of formation, remaining unaltered over cosmic time.} 
\label{figure:4}
\end{figure*}

\section{Summary}

In this work we have studied a sample of 8 compact massive galaxies selected from the \textsc{hetmgs} that are candidates for hosting \"UMBHs and that are therefore considered extreme outliers from the correlation that the SMBHs and the mass of their host galaxies follow. We propose that they deviate because the host galaxies did not structurally evolve in the way expected for massive galaxies due to their relic nature. Our main results are here briefly summarized:

\begin{itemize}

\item We find that 7 out of the 8 \"UMBH host candidates fulfill the criteria to be considered a relic galaxy (i.e. R$_{\mathrm{e}}$$<$2
kpc and t$>$10 Gyr), while one of them is a dubious case. Therefore, selecting galaxies that are extreme outliers in the SMBH mass
scaling relations could represent a new way to find the elusive relic galaxies in the nearby Universe.

\item When we plot the M$_{\bullet}$-$\sigma$ and M$_{\bullet}$-M$_\mathrm{bulge}$ relations, the loci of the galaxies hosting \"UMBHs should represent the position of the massive galaxy population at z$\sim$2. As cosmic time progresses, galaxies following normal growth paths should evolve towards galaxies with larger host masses in those planes.
\item We can pose a lower limit to the age on the SMBH growth at $\sim$\,10\,Gyr, based on the derived SFHs.
\end{itemize}

\acknowledgments
This work was supported by the Japan Society for the Promotion of Science (JSPS) Grant-in-Aid for
Scientific Research (KAKENHI) Number 23224005 and by the Spanish Ministerio de Econom\'ia y
Competitividad (MINECO; grants AYA2009-11137, AYA2011-25527 and AYA2013-48226-C3-1-P). 

\appendix
\section{{\tt STARLIGHT} and the Star Formation Histories}
We briefly summarize here how the full-spectral-fitting code {\tt STARLIGHT} works, but the reader is referred to the source code papers for a more detailed explanation (\citealt{CidFernandes2005}, \citealt{CidFernandes2010}). The code models the extinction as due to foreground dust, and different reddening-laws can be selected to correct from Galactic extinction. Then, it finds the fraction $x_{j}$ that a given $j$th SSP contributes to the total flux of the galaxy (normalized to a certain wavelength $\lambda_{0}$), creating a synthetic spectrum. In other words, it creates a combination of SSPs that best resemble the observed galaxy spectrum and that minimize the $\chi^{2}$. The first two parameters presented in Table 2 are indicative of the quality of the fit.  [$\chi^{2}$] represents the total $\chi^{2}$ divided by the number of $\lambda$'s used in the fit. The second parameter, \textit{adev}, is a proxy for the mean deviation over all fitted pixels. Good values for a fit are those with \textit{adev} below 2-3$\%$. A visual inspection of the fits and their residuals confirms the quality of the results obtained (see Figures A1 and A2).\\
From each $x_{j}$ we can derive its contribution in mass, $\gamma_{j}$, directly from the mass-to-light ratio of each individual SSP (from the models), to obtain the desired mass-weighted estimates such as the star formation episodes and mean stellar populations parameters, which are derived as:

\begin{equation}
\langle t_{\star} \rangle _{M}\,=\,\sum_{j=1}^{N_{\star}}\gamma_{j}\,t_{j}\,; \mathrm{\,\,for\,\,the\,\,age}                                                                                                                                                                                                                                                                                                                                                                                                                                                                                                                                                                                                                                                                                                                                                                                                                                                                                                                                                                                                                                                                                                      \end{equation} \\
\begin{equation}
\langle Z_{\star} \rangle _{M}\,=\,\sum_{j=1}^{N_{\star}}\gamma_{j}Z_{j}\,; \mathrm{\,\,for\,\,the\,\,metallicity}                                                                                                                                                                                                                                                                                                                                                                                                                                                                                                                                                                                                                                                                                                                                                                                                                                                                                                                                                                                                                                                                                                         \end{equation}
\\
We can therefore reconstruct the SFH of the galaxy by summing up all the fractions of stellar mass created at a given epoch, as shown in Figures A1 and A2. They present the whole spectroscopic sample data and their derived SFHs. Like in Figure 3, the left panels show the galaxy spectrum with fit from {\tt STARLIGHT} and the right side ones, the derived SFHs in terms of stellar mass. 

 \begin{figure*}
\centering
\includegraphics[scale=0.7]{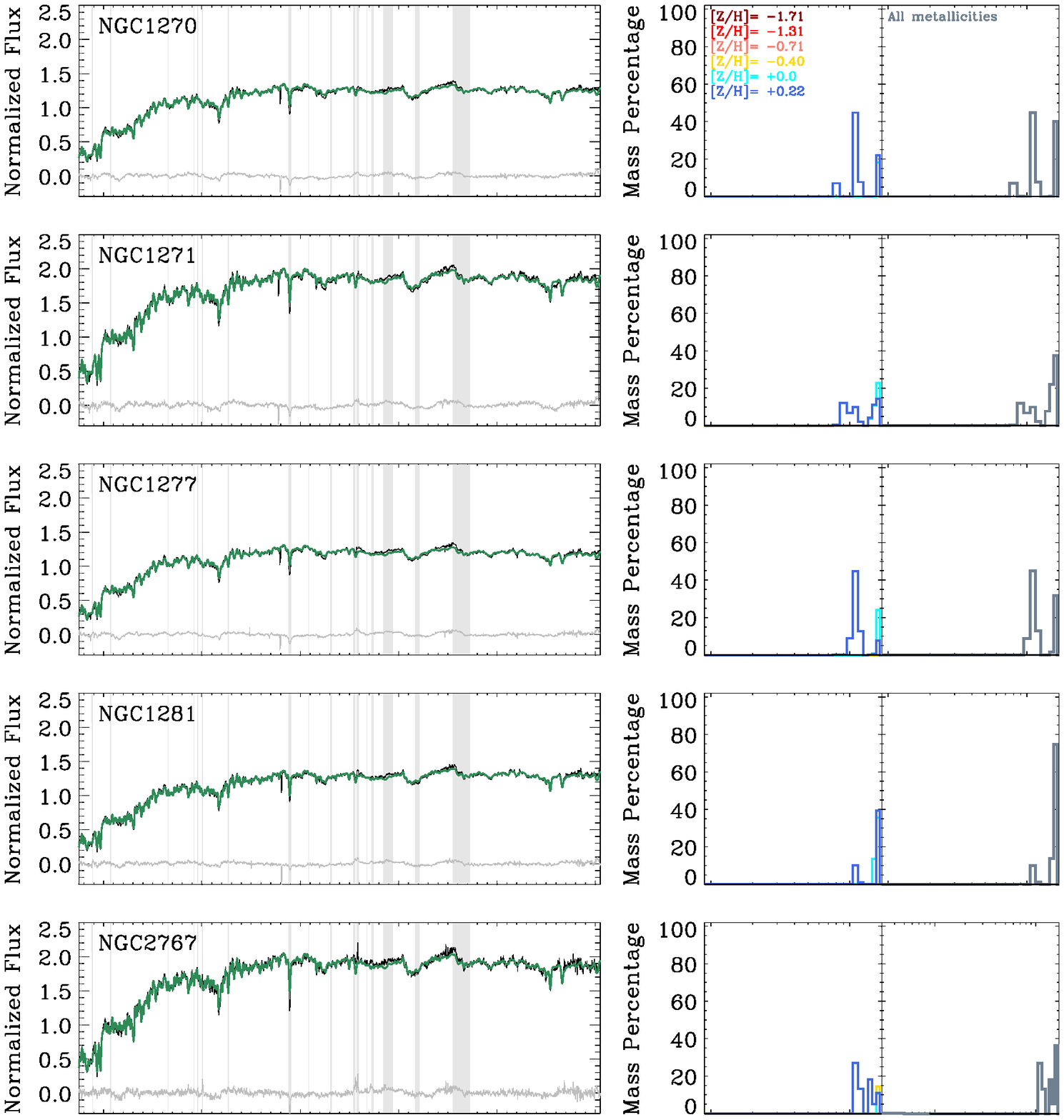}\\
\includegraphics[scale=0.7]{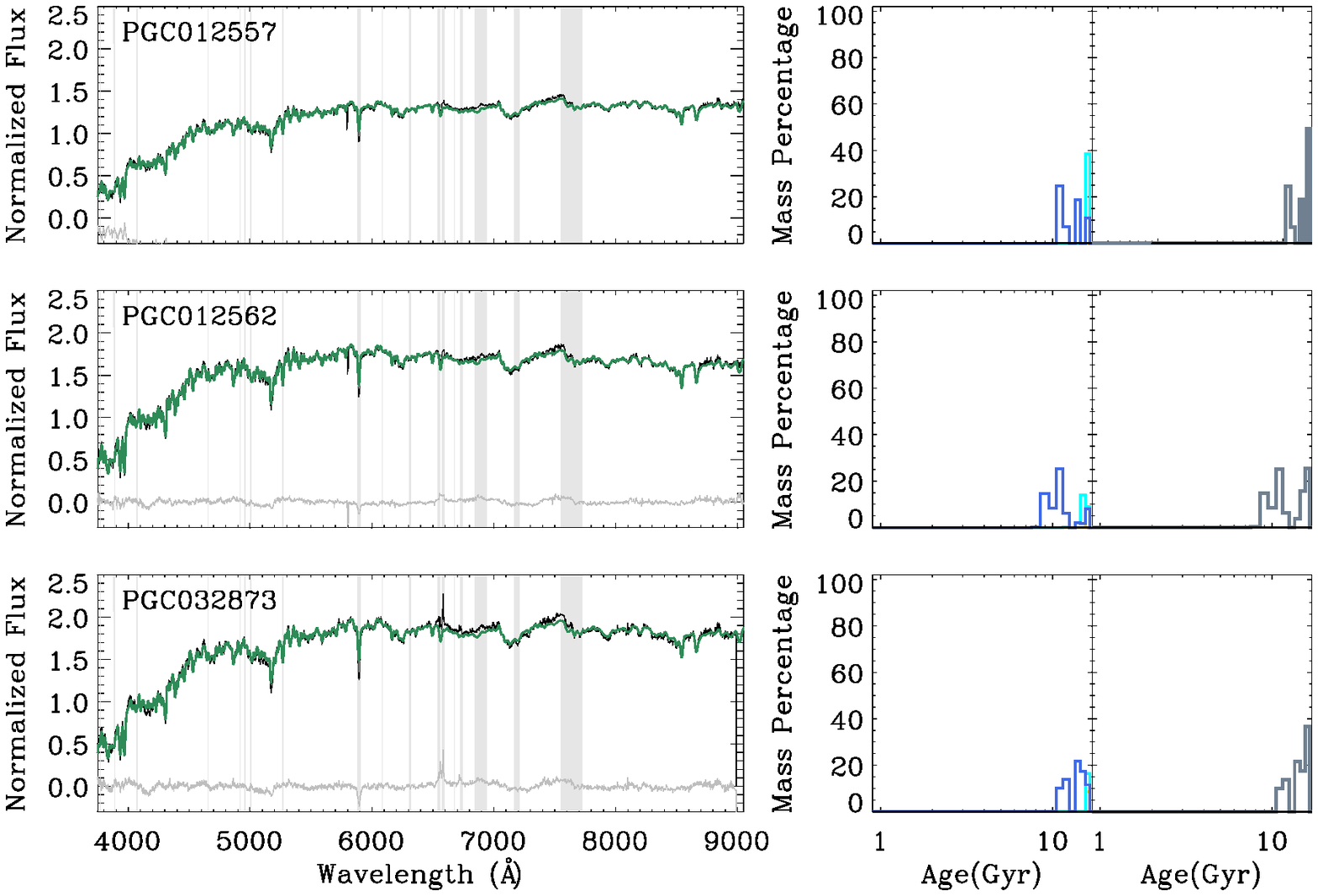}\\
\caption{Star formation histories derived for the sample of \"UMBH host candidates. Each row corresponds to a  galaxy. Left panel shows the galaxy spectrum (black), the fitting (green), and the residuals from the fitting and the masked spectral regions to avoid possible emission line contributions (grey). The right panels show the mass-weighted SFHs considering both the contribution from each individual metallicity and from the total metallicity.}
\label{figure:A1}
\end{figure*}

 \begin{figure*}
\centering
\includegraphics[scale=0.7]{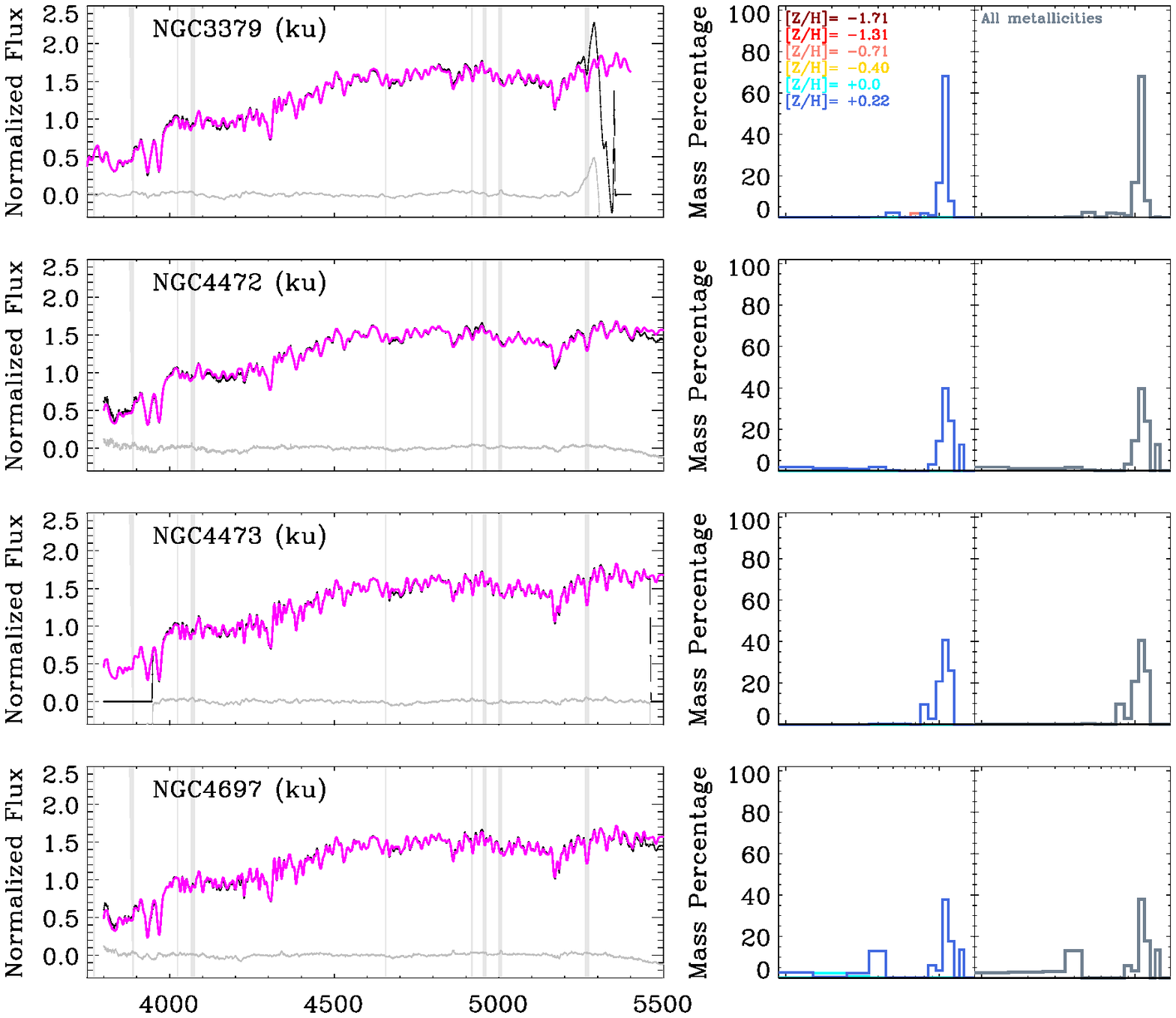}\\
 \includegraphics[scale=0.7]{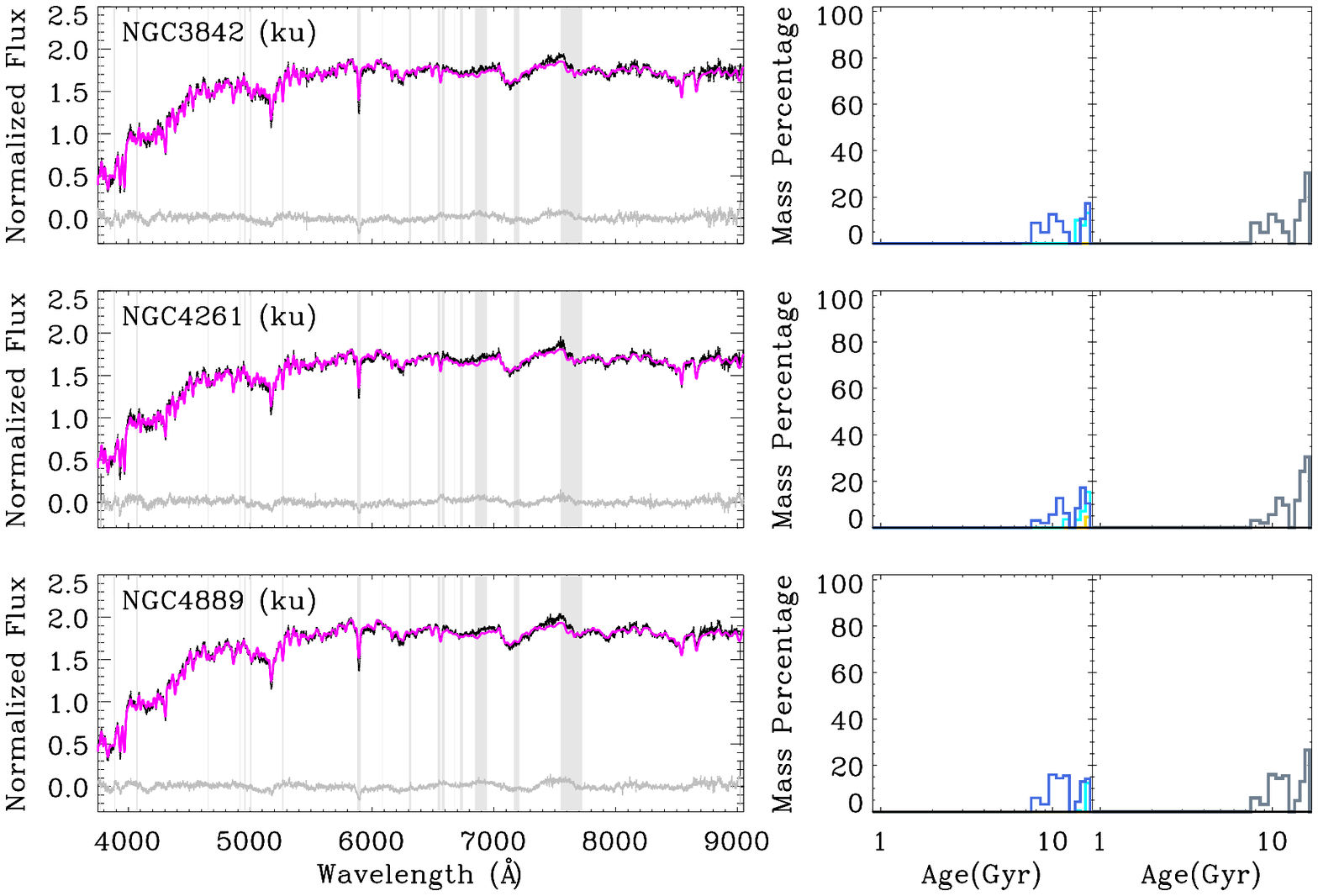}\\
\caption{As in Figure A1 but now for the control sample.}
\label{figure:A2}
\end{figure*}

%\bibliography{biblio_SMBHS_sdss2}

\begin{thebibliography}{72}
\expandafter\ifx\csname natexlab\endcsname\relax\def\natexlab#1{#1}\fi

\bibitem[{{Beifiori} {et~al}\mbox{.}(2012){Beifiori}, {Courteau}, {Corsini}, \&
  {Zhu}}]{Beifiori2012}
{Beifiori} A., {Courteau} S., {Corsini} E.~M., {Zhu} Y., 2012, \mnras, 419,
  2497

\bibitem[{{Bogd{\'a}n} {et~al}\mbox{.}(2012){Bogd{\'a}n}, {Forman},
  {Zhuravleva}, {Mihos}, {Kraft}, {Harding}, {Guo}, {Li}, {Churazov},
  {Vikhlinin}, {Nulsen}, {Schindler}, \& {Jones}}]{Bogdan2012}
{Bogd{\'a}n} {\'A}. {et~al.}, 2012, \apj, 753, 140

\bibitem[{{Buitrago} {et~al}\mbox{.}(2008){Buitrago}, {Trujillo}, {Conselice},
  {Bouwens}, {Dickinson}, \& {Yan}}]{Buitrago2008}
{Buitrago} F., {Trujillo} I., {Conselice} C.~J., {Bouwens} R.~J., {Dickinson}
  M., {Yan} H., 2008, \apjl, 687, L61

\bibitem[{{Cappellari} {et~al}\mbox{.}(2011){Cappellari}, {Emsellem},
  {Krajnovi{\'c}}, {McDermid}, {Scott}, {Verdoes Kleijn}, {Young}, {Alatalo},
  {Bacon}, {Blitz}, {Bois}, {Bournaud}, {Bureau}, {Davies}, {Davis}, {de
  Zeeuw}, {Duc}, {Khochfar}, {Kuntschner}, {Lablanche}, {Morganti}, {Naab},
  {Oosterloo}, {Sarzi}, {Serra}, \& {Weijmans}}]{Cappellari2011}
{Cappellari} M. {et~al.}, 2011, \mnras, 413, 813

\bibitem[{{Cappellari} {et~al}\mbox{.}(2012){Cappellari}, {McDermid},
  {Alatalo}, {Blitz}, {Bois}, {Bournaud}, {Bureau}, {Crocker}, {Davies},
  {Davis}, {de Zeeuw}, {Duc}, {Emsellem}, {Khochfar}, {Krajnovi{\'c}},
  {Kuntschner}, {Lablanche}, {Morganti}, {Naab}, {Oosterloo}, {Sarzi}, {Scott},
  {Serra}, {Weijmans}, \& {Young}}]{Cappellari2012}
---, 2012, \nat, 484, 485

\bibitem[{{Carrasco}, {Conselice} \& {Trujillo}(2010){Carrasco}, {Conselice},
  \& {Trujillo}}]{Carrasco2010}
{Carrasco} E.~R., {Conselice} C.~J., {Trujillo} I., 2010, \mnras, 405, 2253

\bibitem[{{Cenarro} \& {Trujillo}(2009)}]{Cenarro2009}
{Cenarro} A.~J., {Trujillo} I., 2009, \apjl, 696, L43

\bibitem[{{Cid Fernandes} \& {Gonz{\'a}lez Delgado}(2010)}]{CidFernandes2010}
{Cid Fernandes} R., {Gonz{\'a}lez Delgado} R.~M., 2010, \mnras, 403, 780

\bibitem[{{Cid Fernandes} {et~al}\mbox{.}(2005){Cid Fernandes}, {Mateus},
  {Sodr{\'e}}, {Stasi{\'n}ska}, \& {Gomes}}]{CidFernandes2005}
{Cid Fernandes} R., {Mateus} A., {Sodr{\'e}} L., {Stasi{\'n}ska} G., {Gomes}
  J.~M., 2005, \mnras, 358, 363

\bibitem[{{Conroy} \& {van Dokkum}(2012)}]{Conroy2012a}
{Conroy} C., {van Dokkum} P., 2012, \apj, 747, 69

\bibitem[{{Emsellem}(2013)}]{Emsellem2013}
{Emsellem} E., 2013, \mnras, 433, 1862

\bibitem[{{Fabian}(1999)}]{Fabian1999}
{Fabian} A.~C., 1999, \mnras, 308, L39

\bibitem[{{Fabian} {et~al}\mbox{.}(2013){Fabian}, {Sanders}, {Haehnelt},
  {Rees}, \& {Miller}}]{Fabian2013}
{Fabian} A.~C., {Sanders} J.~S., {Haehnelt} M., {Rees} M.~J., {Miller} J.~M.,
  2013, \mnras, 431, L38

\bibitem[{{Ferrarese} \& {Merritt}(2000)}]{Ferrarese2000}
{Ferrarese} L., {Merritt} D., 2000, \apjl, 539, L9

\bibitem[{{Ferr{\'e}-Mateu}, {Vazdekis} \& {de la Rosa}(2013){Ferr{\'e}-Mateu},
  {Vazdekis}, \& {de la Rosa}}]{Ferre-Mateu2013}
{Ferr{\'e}-Mateu} A., {Vazdekis} A., {de la Rosa} I.~G., 2013, \mnras, 431, 440

\bibitem[{{Ferr{\'e}-Mateu} {et~al}\mbox{.}(2012){Ferr{\'e}-Mateu}, {Vazdekis},
  {Trujillo}, {S{\'a}nchez-Bl{\'a}zquez}, {Ricciardelli}, \& {de la
  Rosa}}]{Ferre-Mateu2012}
{Ferr{\'e}-Mateu} A., {Vazdekis} A., {Trujillo} I., {S{\'a}nchez-Bl{\'a}zquez}
  P., {Ricciardelli} E., {de la Rosa} I.~G., 2012, \mnras, 423, 632

\bibitem[{{Ferreras} {et~al}\mbox{.}(2013){Ferreras}, {La Barbera}, {de la
  Rosa}, {Vazdekis}, {de Carvalho}, {Falc{\'o}n-Barroso}, \&
  {Ricciardelli}}]{Ferreras2013}
{Ferreras} I., {La Barbera} F., {de la Rosa} I.~G., {Vazdekis} A., {de
  Carvalho} R.~R., {Falc{\'o}n-Barroso} J., {Ricciardelli} E., 2013, \mnras,
  429, L15

\bibitem[{{Gebhardt} {et~al}\mbox{.}(2000){Gebhardt}, {Bender}, {Bower},
  {Dressler}, {Faber}, {Filippenko}, {Green}, {Grillmair}, {Ho}, {Kormendy},
  {Lauer}, {Magorrian}, {Pinkney}, {Richstone}, \& {Tremaine}}]{Gebhardt2000}
{Gebhardt} K. {et~al.}, 2000, \apjl, 539, L13

\bibitem[{{Graham} {et~al}\mbox{.}(2011){Graham}, {Onken}, {Athanassoula}, \&
  {Combes}}]{Graham2011}
{Graham} A.~W., {Onken} C.~A., {Athanassoula} E., {Combes} F., 2011, \mnras,
  412, 2211

\bibitem[{{Graham} \& {Scott}(2013)}]{Graham2013}
{Graham} A.~W., {Scott} N., 2013, \apj, 764, 151

\bibitem[{{Greene}, {Peng} \& {Ludwig}(2010){Greene}, {Peng}, \&
  {Ludwig}}]{Greene2010}
{Greene} J.~E., {Peng} C.~Y., {Ludwig} R.~R., 2010, \apj, 709, 937

\bibitem[{{G{\"u}ltekin} {et~al}\mbox{.}(2009){G{\"u}ltekin}, {Richstone},
  {Gebhardt}, {Lauer}, {Tremaine}, {Aller}, {Bender}, {Dressler}, {Faber},
  {Filippenko}, {Green}, {Ho}, {Kormendy}, {Magorrian}, {Pinkney}, \&
  {Siopis}}]{Gueltekin2009}
{G{\"u}ltekin} K. {et~al.}, 2009, \apj, 698, 198

\bibitem[{{H{\"a}ring} \& {Rix}(2004)}]{Haering2004}
{H{\"a}ring} N., {Rix} H.-W., 2004, \apjl, 604, L89

\bibitem[{{Hilz}, {Naab} \& {Ostriker}(2013){Hilz}, {Naab}, \&
  {Ostriker}}]{Hilz2013}
{Hilz} M., {Naab} T., {Ostriker} J.~P., 2013, \mnras, 429, 2924

\bibitem[{{Hilz} {et~al}\mbox{.}(2012){Hilz}, {Naab}, {Ostriker}, {Thomas},
  {Burkert}, \& {Jesseit}}]{Hilz2012}
{Hilz} M., {Naab} T., {Ostriker} J.~P., {Thomas} J., {Burkert} A., {Jesseit}
  R., 2012, \mnras, 425, 3119

\bibitem[{{Hopkins} {et~al}\mbox{.}(2006){Hopkins}, {Hernquist}, {Cox},
  {Robertson}, \& {Springel}}]{Hopkins2006}
{Hopkins} P.~F., {Hernquist} L., {Cox} T.~J., {Robertson} B., {Springel} V.,
  2006, \apjs, 163, 50

\bibitem[{{Jahnke} {et~al}\mbox{.}(2009){Jahnke}, {Bongiorno}, {Brusa},
  {Capak}, {Cappelluti}, {Cisternas}, {Civano}, {Colbert}, {Comastri}, {Elvis},
  {Hasinger}, {Ilbert}, {Impey}, {Inskip}, {Koekemoer}, {Lilly}, {Maier},
  {Merloni}, {Riechers}, {Salvato}, {Schinnerer}, {Scoville}, {Silverman},
  {Taniguchi}, {Trump}, \& {Yan}}]{Jahnke2009}
{Jahnke} K. {et~al.}, 2009, \apjl, 706, L215

\bibitem[{{Jahnke} \& {Macci{\`o}}(2011)}]{Jahnke2011}
{Jahnke} K., {Macci{\`o}} A.~V., 2011, \apj, 734, 92

\bibitem[{{Kauffmann} {et~al}\mbox{.}(2003){Kauffmann}, {Heckman}, {White},
  {Charlot}, {Tremonti}, {Brinchmann}, {Bruzual}, {Peng}, {Seibert},
  {Bernardi}, {Blanton}, {Brinkmann}, {Castander}, {Cs{\'a}bai}, {Fukugita},
  {Ivezic}, {Munn}, {Nichol}, {Padmanabhan}, {Thakar}, {Weinberg}, \&
  {York}}]{Kauffmann2003}
{Kauffmann} G. {et~al.}, 2003, \mnras, 341, 33

\bibitem[{{Khandai} {et~al}\mbox{.}(2012){Khandai}, {Feng}, {DeGraf}, {Di
  Matteo}, \& {Croft}}]{Khandai2012}
{Khandai} N., {Feng} Y., {DeGraf} C., {Di Matteo} T., {Croft} R.~A.~C., 2012,
  \mnras, 423, 2397

\bibitem[{{Kormendy} \& {Ho}(2013)}]{Kormendy2013}
{Kormendy} J., {Ho} L.~C., 2013, \araa, 51, 511

\bibitem[{{Kormendy} \& {Richstone}(1995)}]{Kormendy1995}
{Kormendy} J., {Richstone} D., 1995, \araa, 33, 581

\bibitem[{{Krajnovi{\'c}} {et~al}\mbox{.}(2013){Krajnovi{\'c}}, {Alatalo},
  {Blitz}, {Bois}, {Bournaud}, {Bureau}, {Cappellari}, {Davies}, {Davis}, {de
  Zeeuw}, {Duc}, {Emsellem}, {Khochfar}, {Kuntschner}, {McDermid}, {Morganti},
  {Naab}, {Oosterloo}, {Sarzi}, {Scott}, {Serra}, {Weijmans}, \&
  {Young}}]{Krajnovic2013}
{Krajnovi{\'c}} D. {et~al.}, 2013, \mnras, 432, 1768

\bibitem[{{Kulier} {et~al}\mbox{.}(2015){Kulier}, {Ostriker}, {Natarajan},
  {Lackner}, \& {Cen}}]{Kulier2015}
{Kulier} A., {Ostriker} J.~P., {Natarajan} P., {Lackner} C.~N., {Cen} R., 2015,
  \apj, 799, 178

\bibitem[{{La Barbera} {et~al}\mbox{.}(2013){La Barbera}, {Ferreras},
  {Vazdekis}, {de la Rosa}, {de Carvalho}, {Trevisan}, {Falc{\'o}n-Barroso}, \&
  {Ricciardelli}}]{LaBarbera2013}
{La Barbera} F., {Ferreras} I., {Vazdekis} A., {de la Rosa} I.~G., {de
  Carvalho} R.~R., {Trevisan} M., {Falc{\'o}n-Barroso} J., {Ricciardelli} E.,
  2013, \mnras, 433, 3017

\bibitem[{{L{\"a}sker} {et~al}\mbox{.}(2013){L{\"a}sker}, {van den Bosch}, {van
  de Ven}, {Ferreras}, {La Barbera}, {Vazdekis}, \&
  {Falc{\'o}n-Barroso}}]{Laesker2013}
{L{\"a}sker} R., {van den Bosch} R.~C.~E., {van de Ven} G., {Ferreras} I., {La
  Barbera} F., {Vazdekis} A., {Falc{\'o}n-Barroso} J., 2013, \mnras, 434, L31

\bibitem[{{Magorrian} {et~al}\mbox{.}(1998){Magorrian}, {Tremaine},
  {Richstone}, {Bender}, {Bower}, {Dressler}, {Faber}, {Gebhardt}, {Green},
  {Grillmair}, {Kormendy}, \& {Lauer}}]{Magorrian1998}
{Magorrian} J. {et~al.}, 1998, \aj, 115, 2285

\bibitem[{{Mart{\'{\i}}n-Navarro} {et~al}\mbox{.}(2015){Mart{\'{\i}}n-Navarro},
  {La Barbera}, {Vazdekis}, {Ferr{\'e}-Mateu}, {Trujillo}, \&
  {Beasley}}]{Martin-Navarro2015}
{Mart{\'{\i}}n-Navarro} I., {La Barbera} F., {Vazdekis} A., {Ferr{\'e}-Mateu}
  A., {Trujillo} I., {Beasley} M.~A., 2015, ArXiv e-prints:1505.01485

\bibitem[{{McConnell} \& {Ma}(2013)}]{McConnell2013}
{McConnell} N.~J., {Ma} C.-P., 2013, \apj, 764, 184

\bibitem[{{Menci} {et~al}\mbox{.}(2008){Menci}, {Fiore}, {Puccetti}, \&
  {Cavaliere}}]{Menci2008}
{Menci} N., {Fiore} F., {Puccetti} S., {Cavaliere} A., 2008, \apj, 686, 219

\bibitem[{{Merloni} {et~al}\mbox{.}(2010){Merloni}, {Bongiorno}, {Bolzonella},
  {Brusa}, {Civano}, {Comastri}, {Elvis}, {Fiore}, {Gilli}, {Hao}, {Jahnke},
  {Koekemoer}, {Lusso}, {Mainieri}, {Mignoli}, {Miyaji}, {Renzini}, {Salvato},
  {Silverman}, {Trump}, {Vignali}, {Zamorani}, {Capak}, {Lilly}, {Sanders},
  {Taniguchi}, {Bardelli}, {Carollo}, {Caputi}, {Contini}, {Coppa}, {Cucciati},
  {de la Torre}, {de Ravel}, {Franzetti}, {Garilli}, {Hasinger}, {Impey},
  {Iovino}, {Iwasawa}, {Kampczyk}, {Kneib}, {Knobel}, {Kova{\v c}},
  {Lamareille}, {Le Borgne}, {Le Brun}, {Le F{\`e}vre}, {Maier}, {Pello},
  {Peng}, {Perez Montero}, {Ricciardelli}, {Scodeggio}, {Tanaka}, {Tasca},
  {Tresse}, {Vergani}, \& {Zucca}}]{Merloni2010}
{Merloni} A. {et~al.}, 2010, \apj, 708, 137

\bibitem[{{Naab}, {Johansson} \& {Ostriker}(2009){Naab}, {Johansson}, \&
  {Ostriker}}]{Naab2009}
{Naab} T., {Johansson} P.~H., {Ostriker} J.~P., 2009, \apjl, 699, L178

\bibitem[{{Oogi} \& {Habe}(2013)}]{Oogi2013}
{Oogi} T., {Habe} A., 2013, \mnras, 428, 641

\bibitem[{{Oser} {et~al}\mbox{.}(2012){Oser}, {Naab}, {Ostriker}, \&
  {Johansson}}]{Oser2012}
{Oser} L., {Naab} T., {Ostriker} J.~P., {Johansson} P.~H., 2012, \apj, 744, 63

\bibitem[{{Oser} {et~al}\mbox{.}(2010){Oser}, {Ostriker}, {Naab}, {Johansson},
  \& {Burkert}}]{Oser2010}
{Oser} L., {Ostriker} J.~P., {Naab} T., {Johansson} P.~H., {Burkert} A., 2010,
  \apj, 725, 2312

\bibitem[{{Petri}, {Ferrara} \& {Salvaterra}(2012){Petri}, {Ferrara}, \&
  {Salvaterra}}]{Petri2012}
{Petri} A., {Ferrara} A., {Salvaterra} R., 2012, \mnras, 422, 1690

\bibitem[{{Poggianti} {et~al}\mbox{.}(2013){Poggianti}, {Calvi}, {Bindoni},
  {D'Onofrio}, {Moretti}, {Valentinuzzi}, {Fasano}, {Fritz}, {De Lucia},
  {Vulcani}, {Bettoni}, {Gullieuszik}, \& {Omizzolo}}]{Poggianti2013}
{Poggianti} B.~M. {et~al.}, 2013, \apj, 762, 77

\bibitem[{{Quilis} \& {Trujillo}(2013)}]{Quilis2013}
{Quilis} V., {Trujillo} I., 2013, \apjl, 773, L8

\bibitem[{{Reines} {et~al}\mbox{.}(2011){Reines}, {Sivakoff}, {Johnson}, \&
  {Brogan}}]{Reines2011}
{Reines} A.~E., {Sivakoff} G.~R., {Johnson} K.~E., {Brogan} C.~L., 2011, \nat,
  470, 66

\bibitem[{{Ricciardelli} {et~al}\mbox{.}(2012){Ricciardelli}, {Vazdekis},
  {Cenarro}, \& {Falc{\'o}n-Barroso}}]{Ricciardelli2012}
{Ricciardelli} E., {Vazdekis} A., {Cenarro} A.~J., {Falc{\'o}n-Barroso} J.,
  2012, \mnras, 424, 172

\bibitem[{{Rusli} {et~al}\mbox{.}(2011){Rusli}, {Thomas}, {Erwin}, {Saglia},
  {Nowak}, \& {Bender}}]{Rusli2011}
{Rusli} S.~P., {Thomas} J., {Erwin} P., {Saglia} R.~P., {Nowak} N., {Bender}
  R., 2011, \mnras, 410, 1223

\bibitem[{{Saulder}, {van den Bosch} \& {Mieske}(2015){Saulder}, {van den
  Bosch}, \& {Mieske}}]{Saulder2015}
{Saulder} C., {van den Bosch} R.~C.~E., {Mieske} S., 2015, ArXiv e-prints:
  1503.05117

\bibitem[{{Schulze} \& {Gebhardt}(2011)}]{Schulze2011}
{Schulze} A., {Gebhardt} K., 2011, \apj, 729, 21

\bibitem[{{Shen} {et~al}\mbox{.}(2003){Shen}, {Mo}, {White}, {Blanton},
  {Kauffmann}, {Voges}, {Brinkmann}, \& {Csabai}}]{Shen2003}
{Shen} S., {Mo} H.~J., {White} S.~D.~M., {Blanton} M.~R., {Kauffmann} G.,
  {Voges} W., {Brinkmann} J., {Csabai} I., 2003, \mnras, 343, 978

\bibitem[{{Somerville} {et~al}\mbox{.}(2008){Somerville}, {Hopkins}, {Cox},
  {Robertson}, \& {Hernquist}}]{Somerville2008}
{Somerville} R.~S., {Hopkins} P.~F., {Cox} T.~J., {Robertson} B.~E.,
  {Hernquist} L., 2008, \mnras, 391, 481

\bibitem[{{Spiniello} {et~al}\mbox{.}(2012){Spiniello}, {Trager}, {Koopmans},
  \& {Chen}}]{Spiniello2012}
{Spiniello} C., {Trager} S.~C., {Koopmans} L.~V.~E., {Chen} Y.~P., 2012, \apjl,
  753, L32

\bibitem[{{Taylor} {et~al}\mbox{.}(2010){Taylor}, {Franx}, {Glazebrook},
  {Brinchmann}, {van der Wel}, \& {van Dokkum}}]{Taylor2010}
{Taylor} E.~N., {Franx} M., {Glazebrook} K., {Brinchmann} J., {van der Wel} A.,
  {van Dokkum} P.~G., 2010, \apj, 720, 723

\bibitem[{{Trujillo}, {Carrasco} \& {Ferr{\'e}-Mateu}(2012){Trujillo},
  {Carrasco}, \& {Ferr{\'e}-Mateu}}]{Trujillo2012}
{Trujillo} I., {Carrasco} E.~R., {Ferr{\'e}-Mateu} A., 2012, \apj, 751, 45

\bibitem[{{Trujillo} {et~al}\mbox{.}(2009){Trujillo}, {Cenarro}, {de
  Lorenzo-C{\'a}ceres}, {Vazdekis}, {de la Rosa}, \& {Cava}}]{Trujillo2009}
{Trujillo} I., {Cenarro} A.~J., {de Lorenzo-C{\'a}ceres} A., {Vazdekis} A., {de
  la Rosa} I.~G., {Cava} A., 2009, \apjl, 692, L118

\bibitem[{{Trujillo} {et~al}\mbox{.}(2007){Trujillo}, {Conselice}, {Bundy},
  {Cooper}, {Eisenhardt}, \& {Ellis}}]{Trujillo2007}
{Trujillo} I., {Conselice} C.~J., {Bundy} K., {Cooper} M.~C., {Eisenhardt} P.,
  {Ellis} R.~S., 2007, \mnras, 382, 109

\bibitem[{{Trujillo} {et~al}\mbox{.}(2014){Trujillo}, {Ferr{\'e}-Mateu},
  {Balcells}, {Vazdekis}, \& {S{\'a}nchez-Bl{\'a}zquez}}]{Trujillo2014}
{Trujillo} I., {Ferr{\'e}-Mateu} A., {Balcells} M., {Vazdekis} A.,
  {S{\'a}nchez-Bl{\'a}zquez} P., 2014, \apjl, 780, L20

\bibitem[{{Trujillo}, {Ferreras} \& {de La Rosa}(2011){Trujillo}, {Ferreras},
  \& {de La Rosa}}]{Trujillo2011}
{Trujillo} I., {Ferreras} I., {de La Rosa} I.~G., 2011, \mnras, 415, 3903

\bibitem[{{Trujillo} {et~al}\mbox{.}(2004){Trujillo}, {Rudnick}, {Rix},
  {Labb{\'e}}, {Franx}, {Daddi}, {van Dokkum}, {F{\"o}rster Schreiber},
  {Kuijken}, {Moorwood}, {R{\"o}ttgering}, {van der Wel}, {van der Werf}, \&
  {van Starkenburg}}]{Trujillo2004}
{Trujillo} I. {et~al.}, 2004, \apj, 604, 521

\bibitem[{{van den Bosch} {et~al}\mbox{.}(2012){van den Bosch}, {Gebhardt},
  {G{\"u}ltekin}, {van de Ven}, {van der Wel}, \& {Walsh}}]{vandenBosch2012}
{van den Bosch} R.~C.~E., {Gebhardt} K., {G{\"u}ltekin} K., {van de Ven} G.,
  {van der Wel} A., {Walsh} J.~L., 2012, \nat, 491, 729

\bibitem[{{van den Bosch} {et~al}\mbox{.}(2015){van den Bosch}, {Gebhardt},
  {G{\"u}ltekin}, {Y{\i}ld{\i}r{\i}m}, \& {Walsh}}]{vandenBosch2015}
{van den Bosch} R.~C.~E., {Gebhardt} K., {G{\"u}ltekin} K., {Y{\i}ld{\i}r{\i}m}
  A., {Walsh} J.~L., 2015, \apjs, 218, 10

\bibitem[{{Vazdekis} {et~al}\mbox{.}(2012){Vazdekis}, {Ricciardelli},
  {Cenarro}, {Rivero-Gonz{\'a}lez}, {D{\'{\i}}az-Garc{\'{\i}}a}, \&
  {Falc{\'o}n-Barroso}}]{Vazdekis2012}
{Vazdekis} A., {Ricciardelli} E., {Cenarro} A.~J., {Rivero-Gonz{\'a}lez} J.~G.,
  {D{\'{\i}}az-Garc{\'{\i}}a} L.~A., {Falc{\'o}n-Barroso} J., 2012, \mnras,
  424, 157

\bibitem[{{Walsh} {et~al}\mbox{.}(2015 subm.){Walsh}, {van den Bosch},
  {Gebhardt}, Y{\i}ld{\i}r{\i}m, {G{\"u}ltekin}, \& {Richstone}}]{Walsh2015}
{Walsh} J.~L., {van den Bosch} R.~C.~E., {Gebhardt} K., Y{\i}ld{\i}r{\i}m A.,
  {G{\"u}ltekin} K., {Richstone} D.~O., 2015 subm., \apj

\bibitem[{{Walter} {et~al}\mbox{.}(2004){Walter}, {Carilli}, {Bertoldi},
  {Menten}, {Cox}, {Lo}, {Fan}, \& {Strauss}}]{Walter2004}
{Walter} F., {Carilli} C., {Bertoldi} F., {Menten} K., {Cox} P., {Lo} K.~Y.,
  {Fan} X., {Strauss} M.~A., 2004, \apjl, 615, L17

\bibitem[{{Wellons} {et~al}\mbox{.}(2015){Wellons}, {Torrey}, {Ma},
  {Rodriguez-Gomez}, {Vogelsberger}, {Kriek}, {van Dokkum}, {Nelson}, {Genel},
  {Pillepich}, {Springel}, {Sijacki}, {Snyder}, {Nelson}, {Sales}, \&
  {Hernquist}}]{Wellons2015}
{Wellons} S. {et~al.}, 2015, \mnras, 449, 361

\bibitem[{{Yamada} {et~al}\mbox{.}(2006){Yamada}, {Arimoto}, {Vazdekis}, \&
  {Peletier}}]{Yamada2006}
{Yamada} Y., {Arimoto} N., {Vazdekis} A., {Peletier} R.~F., 2006, \apj, 637,
  200

\bibitem[{Y{\i}ld{\i}r{\i}m {et~al}\mbox{.}(2015 subm.)Y{\i}ld{\i}r{\i}m, {van
  den Bosch}, {Gebhardt}, {G{\"u}ltekin}, {van de Ven}, \&
  {Walsh}}]{Yildirim2015}
Y{\i}ld{\i}r{\i}m A., {van den Bosch} R.~C.~E., {Gebhardt} K., {G{\"u}ltekin}
  K., {van de Ven} G., {Walsh} J.~L., 2015 subm., MNRAS

\bibitem[{{Yoo} {et~al}\mbox{.}(2007){Yoo}, {Miralda-Escud{\'e}}, {Weinberg},
  {Zheng}, \& {Morgan}}]{Yoo2007}
{Yoo} J., {Miralda-Escud{\'e}} J., {Weinberg} D.~H., {Zheng} Z., {Morgan}
  C.~W., 2007, \apj, 667, 813

\end{thebibliography}
%\bibliographystyle{mn2e}

\end{document}